\documentclass[
reprint,
amsmath,amssymb
]{revtex4-2}

\usepackage{graphicx}
\usepackage{dcolumn}
\usepackage{bm}
\usepackage{color}
\usepackage{comment}

\usepackage{hyperref}
\usepackage{txfonts}
\hypersetup{
colorlinks=true,
hidelinks
}

\newcommand{\URL}[1]{{\color{blue}#1}}

\begin{document}

\title{Proper Orthogonal Decomposition of a Superfluid Turbulent Wake}

\author{Sota Yoneda$^{1}$}
\email{yonedaomu@gmail.com}
\author{Hiromitsu Takeuchi$^{1,2}$}
\email{takeuchi@omu.ac.jp}
\affiliation{
$^1$ Department of Physics, Osaka Metropolitan University, 3-3-138 Sugimoto, Osaka 558-8585, Japan \\
$^2$ Department of Physics, Nambu Yoichiro Institute of Theoretical and Experimental Physics (NITEP), Osaka Metropolitan University, 3-3-138 Sugimoto, Osaka 558-8585, Japan
}

\date{\today}

%%%%%%%%%%%%%%%%%%%%%%%%%%%%%%%%%%%%%%%%%%%%%%%%%%%%%%%%%%%%%%%%%%%%%%%%%%%%%%%%%%%%%%%%%%%%%%%%%%%%%%%%%%%%%%%%

\begin{abstract}
Superfluid turbulent wakes behind a square prism are studied theoretically and numerically by proper orthogonal decomposition (POD).
POD is a data science approach that can efficiently extract the principal vibration modes of a physical system, and is widely used in hydrodynamics, including applications in wake structure analysis.
It is not straightforward to apply the conventional POD method to superfluid wake systems, as the superfluid velocity field diverges at the center of a vortex whose circulation is quantized.
We successfully established a POD method by applying appropriate blurring to the vorticity distribution in a two-dimensional superfluid wake.
It is shown that a coherent structure corresponding to two parallel arrays of alternating quantum vortex bundles, called the \textit{quasi-classical} Kármán vortex street, is latent as a distinctive major mode in the superfluid turbulent wakes that were naively thought to be \textit{irregular}.
Since our method is also effective for fluid density, it can be applied to the experimental data analysis for ultra-cold atomic gases.
\end{abstract}

\maketitle

Proper orthogonal decomposition (POD) is a statistical framework capable of efficiently identifying latent structures in large-scale datasets by extracting dominant spatiotemporal modes, and is used in a wide range of fields such as neuroscience \cite{Neuroscience1999}, atmospheric science\cite{Atmospheric2006}, and paleontology \cite{paleontology2014}.
POD was introduced into hydrodynamics by Lumley~\cite{Lumley1967POD}.
Recently, POD has become widely used because the advancement of technology in experiments and numerical simulations has made it easier to obtain large-scale data of time-evolving flow fields.
A representative example is the wake flow formed behind an obstacle.
Since the wake flow significantly influences the drag force acting on the obstacle and is also of engineering significance, there are many cases of POD applications~\cite{Rehimi2008POD,Wang2014POD,Samani2015POD}.
Especially, POD is effective for extracting the coherent structures of Kármán vortex street (KVS), realized in a range of Reynolds number~\cite{Reynolds1883} $10 \lesssim Re \lesssim 10^2$~\cite{Williamson1996KVS}, from the major modes reflecting the periodicity and antisymmetry of the wake \cite{Vitkovicova2020POD,Dewanshu2022POD}.

\begin{figure}[tbp]
\centering
\includegraphics[width=8.5cm]{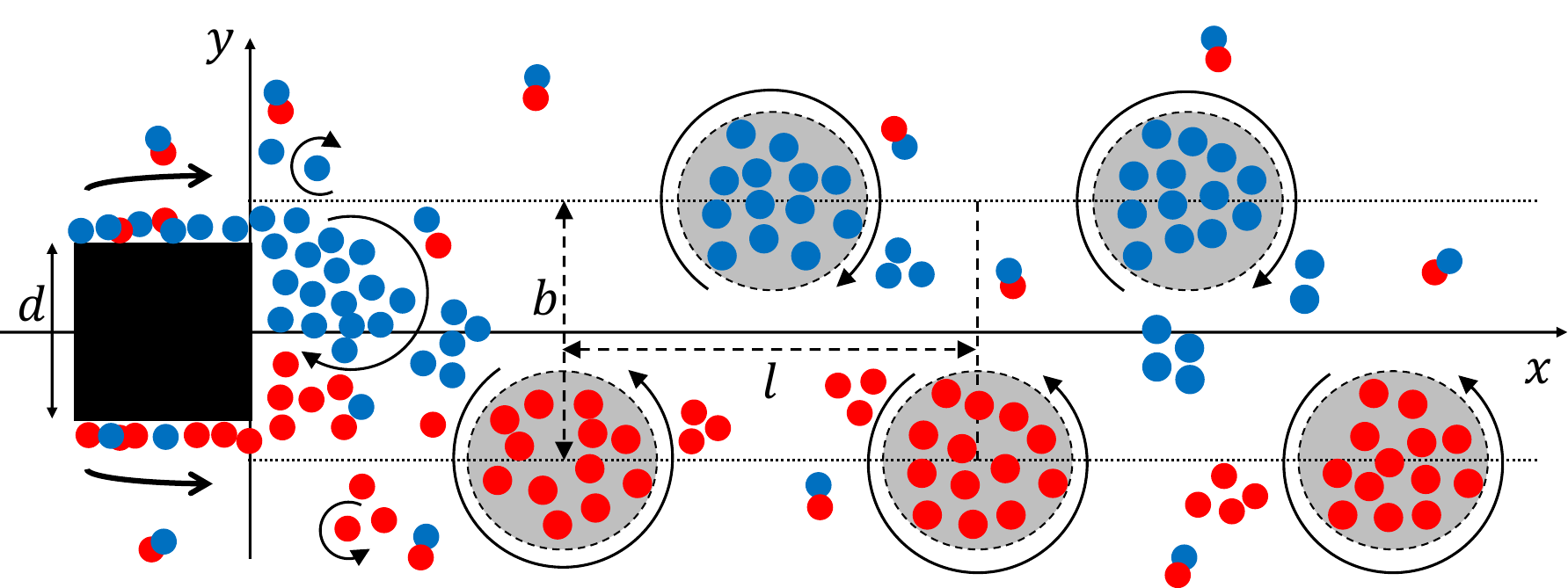}
\caption{
(Color online) Schematic of quasi-classical KVS in a superfluid wake behind a square prisme of size $d$.
Filled circles of two different colors represent quantum vortices of circulations $\kappa$ (red) and $-\kappa$ (blue).
Vortex bundles (dashed circles) form alternating double arrays of period $l$ and spacing $b$.
}
\label{fig:bundlewake}
\end{figure}

Recently, the existence of the KVS has also been reported in superfluid wake numerically~\cite{Sasaki2010wake} and experimentally~\cite{Kwon2016QKarman}.
Superfluids, realized in quantum fluids such as liquid helium and ultracold atomic gases, are inviscid at absolute zero,
and only vortices with elementary circulation $\kappa$, called quantum vortices, are stable as topological defects.
Naively speaking,
the description in terms of the Reynolds number $Re$ or Reynolds similitude would theoretically break down in the inviscid quantum fluids as $Re$ is inversely proportional to the kinematic viscosity $\nu$.
This dilemma can be resolved by introducing the superfluid Reynolds number $Re_s$~\cite{Nore1997bundle,Nore2000Rs,Volovik2003}, which is defined by replacing $\nu$ by a quantity $\sim\kappa$.
This speculation has stimulated many researchers and motivated a series of theoretical and experimental studies on superfluid wake \cite{Reeves2015Rs,Schoepe2015Res,Kwon2016QKarman,Lim2022Res,Schoepe2022Res,Wang2022QKVS,Skrbek2024Res,Takeuchi2024Rs}.
The superfluid Reynolds number was proposed in Ref.~\cite{Volovik2003} based on the vortex-bundle conjecture stating that bundles of quantum vortices behave like a classical vortex, reproducing the Kolmogorov law in the turbulence of a quantum fluid, called the \textit{quasi-classical} turbulence \cite{Vinen2002}.
Accordingly, it is expected that a periodic arrangement of quantum vortex bundles, called a \textit {quasi-classical} KVS as shown in FIG.~\ref{fig:bundlewake}, could appear with a wide range of $Re_s$ in a superfluid wake as a quantum counterpart of KVS of classical fluids.

Contrary to this expectation, the KVS has only been found with very small $Re_s \sim 0.1$, and the street consists double arrays of pairs of like-sign quantum vortices, namely the {\it quantized} KVS~\cite{Sasaki2010wake}.
When the superfluid Reynolds number is larger, no periodic structures have been found, leading to the conclusion that the wake is \textit{irregular} over a wide range of parameters in contrast to classical fluids.
Considering the successful application of POD analysis to extract coherent structures from complex flow phenomena in classical systems,
the POD analysis must be helpful to verify the above conjecture of quasi-classical KVS.
However, POD has not yet been applied to superfluid systems, and the methodology remains undeveloped.

This letter reports on the first application of POD analysis to numerical simulations of a superfluid wake in the Gross-Pitaevskii (GP) model.
Our POD analysis reveal that there are the major modes with a coherent structure in the superfluid turbulent wake that was considered \textit{irregular}.
This coherent structure closely resembles the structure in a classical-fluid wake with KVS,
suggesting quasi-classical KVS.
By computing the spatial and temporal periods of the major modes,
our numerical analysis suggests that superfluid wakes with quasi-classical KVS is characterized by an universal function of the superfluid Reynolds number.

\textit{Analysis settings.}
We consider a two-dimensional flow around a square prism of cross-section area $d\times d$ moving with a fixed speed $U$ in the $x$ direction in a superfluid at rest in the laboratory frame.
When an obstacle is moving at a constant velocity through a superfluid,
the dragless flow, known as the d'Alembert's paradox \cite{Batchelor1967}, is realized until $U$ exceeds a critical velocity $U_c$,
above which vortices are generated.
The velocity $U$ must be sufficiently larger than $U_c$ to form a turbulent wake composed of many vortices.
The critical velocity is strongly dependent on the shape and surface condition of obstacles \cite{Kwon2015Uc,Rica2001Uc,Kwak2023Uc}.
For simplicity, we consider an obstacle with smooth hard walls that the superfluid wavefunction $\psi=\sqrt{n}e^{i\theta}$ cannot penetrate interior at all.
Here, $n$ represent the fluid density and the fluid velocity is given by
$\bm{v}=\frac{\hbar}{m}\bm{\nabla}\theta$ with the reduced Planck constant $\hbar$ and the mass $m$ of Bose particles constituting the superfluid.
Cylinders are often used in previous studies as an obstacle \cite{Frisch1992wake,Sasaki2010wake}, %ハードウォールと見なせるもののみciteする
but its critical speed is about $0.3c$ with the sound speed $c$, independent of size \cite{Rica2001Uc,Pham2005Uc}.
Here, we use a simplest geometry, square, with smaller $U_c$ to realize the ideal situation $U \gg U_c$ \cite{Takeuchi2024Rs}.
The critical velocity of a square prism decreases with its size $d$ as has been confirmed for a plate-like obstacle \cite{Kokubo2025Uc}.

Superfluid wakes is analyzed based on the GP model \cite{Gross1961,Pitaevskii1961} in the inertial frame moving with the obstacle. 
The superfluid wavefunction $\psi$ obeys the nonlinear Schrödinger equation $i\hbar\partial_t \psi=(H + i\hbar U\partial_x)\psi$ with the operator
\begin{equation}
    H= -\frac{\hbar^2 \bm{\nabla}^2}{2m}-\mu +gn+V_{\rm ext}
\end{equation}
Here, $\mu$ and $g$ are the chemical potential and the coupling constant proportional to the s-wave scattering length of the atomic interaction, respectively.
Density in bulk is given by $\mu/g$ and the core size of a quantum vortex is characterized by the healing length $\xi=\hbar/\sqrt{m\mu}$.
The external potential $V_{\rm ext}(\bm{x})$ represents the square obstacle whose interior completely excludes the wave function as schematically illustrated as a black square in Fig.~\ref{fig:bundlewake}.

In order to reproduce the non-equilibrium steady state of a superfluid wake within a finite system, it is necessary to place an energy sink near the boundary of the system far enough away from the prism.
Full-scale numerical calculations of such open systems for superfluid wakes were first realized by Reeves et al. \cite{Reeves2015Rs}.
We used a method similar to theirs and the resulting time evolution equation we solve is
\begin{equation}
    \partial_t\psi=\frac{1}{i\hbar}\left[ H+i\hbar U\partial_x \right]\psi-\gamma(\bm{x})H\psi.
\end{equation}
Here, the third term on the right hand side causes energy dissipation in the laboratory frame and the dissipation coefficient $\gamma(\bm{x})$ is nonzero only in the vicinity of the system boundary.
In contrast to previous studies, the Neumann boundary condition is imposed on the boundaries of the system.
This avoids the problem of quantum vortices emitted backward appearing from the front side due to the periodic boundary condition \cite{Reeves2015Rs}.

Numerical simulations are performed by changing parameters $U$ and $d$ within the ranges from $d/\xi = 12$ to $64$ and from $U/c=0.6$ to $0.9$, which realizes {\it irregular} superfluid wakes.
The system size is fixed to be $15d\times 8 d$.
The computational grid size is $\Delta x=\Delta y=0.5\xi$.
\footnote{(Supplemental Material) Details of our numerical settings and movies of our simulations can be accessed online.}

POD analysis of classical fluids usually uses data of velocity field.
However,
since the phase $\theta$ of $\psi$ plays the role of the velocity potential,
the velocity diverges around the phase singularity at the axis of a quantum vortex,
which makes velocity POD difficult in superfluid systems.
Instead of velocity, POD is applied to vorticity $\bm{\omega}=\bm{\nabla}\times \bm{v}=(0,0,\omega)$, represented by the position $\bm{x}_j$ of the $j$th vortex and the circulation quantum $\kappa=2\pi\hbar/m$ as 
$\omega=\sum_j \Gamma_j\delta(\bm{x}-\bm{x}_j)$ with $\Gamma_j=\pm \kappa$.
In addition to the vorticity, we perform the data analysis on density $n$ that is an important observable in experiments of atomic Bose-Einstein condensates \cite{Kwon2016QKarman}.

\textit{Unfiltered POD.} 
Snapshot POD, a POD method proposed by Sirovich \cite{Sirovich1987Snapshot}, is applied to our numerical data.
This method constructs a matrix from multiple time-series snapshots and treats deviations from the time-averaged data as individual vectors.
By solving an eigenvalue problem based on this matrix, it extracts dominant deviations or spatial modes that contribute most to the system's dynamics.
POD was applied to a superfluid wake after a certain relaxation time $t_\textrm{relax}$.
Here, we set $t_\textrm{relax}=32d/U$,
which is long enough to achieve nonequilibrium steady states.
As shown in \cite{Reeves2015Rs}, the time evolution of a superfluid wake has a characteristic periodic time
\begin{equation}\label{Eq_tau}
\tau=f^{-1}=\frac{d}{\mathit{St} U}, 
\end{equation}
determined by the Strouhal number $\mathit{St}$.
To obtain statistically reliable results,
time evolution data up to more than $40\tau$ after $t=t_\textrm{relax}$ were used for our POD analysis.
In order to reflect the microscopic motion of each quantum vortex,
it is necessary to take snapshots with a sufficiently short period: $\tau/40$.
Therefore, we used 1600 snapshots, which is considerably more than the usual POD analysis.
\footnote{(Supplemental Material) Detailed settings of our POD analysis can be accessed online.}

\begin{figure}
\centering
\includegraphics[width=8.5cm]{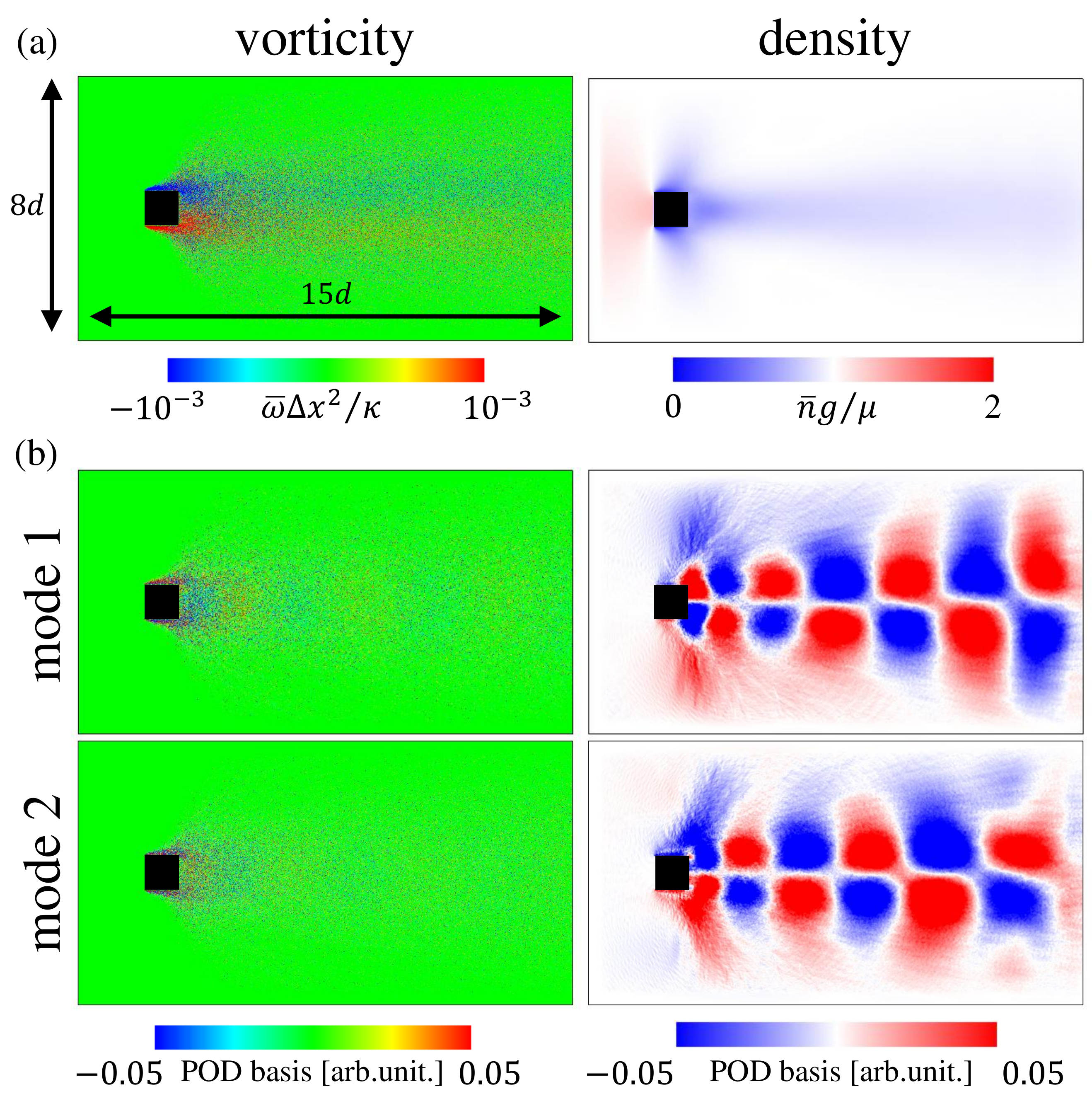}
\caption{
(Color online) Results of the unfiltered POD for vorticity (left) and density (right) for $d=64\xi$, $U=0.7c$, and $Re_s=7.13$.
(a) Normalized mean vorticity $\bar{\omega}\Delta x^2/\kappa$ and density $\bar{n}g/\mu$ with the spatial resolution $\Delta x$ in our numerical simulations. Box size is $15d \times 8d$.
(b)
The first (top) and second (bottom) contributing modes around the mean fields of (a).
}
\label{fig:POD}
\end{figure}

Fig.~\ref{fig:POD} (a) shows a typical example of the time averages of vorticity [$\bar{\omega}(\bm{x})$] and density [$\bar{n}(\bm{x})$] after $t=t_\textrm{relax}$.
Since the fluid must rapidly turn around the forward corner of the square prism,
the local flow velocity is greater there, from which quantum vortices are likely to be generated
\footnote{(Supplemental Material) Videos of the time development of vorticity and density can be accessed online.}.
Vortices with opposite signs of circulation are frequently generated due to the turning flow around the two forward corners.
The effect of this frequent vortex generation on the mean vorticity $\bar{\omega}$ is apparent at the sides of the prism
while vortices are dispersed and their effect is less apparent behind the prism in Fig.~\ref{fig:POD}(a).
In contrast, the mean density $\bar{n}$ clearly shows a groove extending backward.
The appearance of this groove is averaged by the orbits of density holes of radius $\sim\xi$ due to the motion of quantum vortices.
In front of the prism, the fluid is more dense due to compression.
As $U$ increases, shock waves and vortex pairs with opposite circulation are more likely to occur on the sides,
and the mean density become low there.
Many of these vortex pairs either run parallel in the vicinity of the columns or move laterally to reach the energy sink and disappear, so they do not seem to have a significant effect on the wake behind the prism.

For the POD modes with the highest contribution (mode 1) and the second highest one (mode 2),
only the density shows a clear spatial pattern, as shown in Fig.~\ref{fig:POD} (b).
The major modes with antisymmetric and periodic structures appear in pairs.
This fact implies that an antisymmetric wake is a dominant structure
since the modes represent deviations $\delta n=n-\bar{n}$ from the symmetric mean density $\bar{n}$.
Furthermore, the two modes have a half-phase shift in space.
This indicates that the dominant structure propagates in the direction of flow.

\textit{Filtered POD.}
The above result strongly suggests that periodic structures such as KVS are latent in the superfluid turbulent wake.
However, for direct comparison with the KVS of classical fluids, extraction of the major modes is required from the vorticity data including information on the direction of vortex circulation

\begin{figure}
\centering
\includegraphics[width=8.5cm]{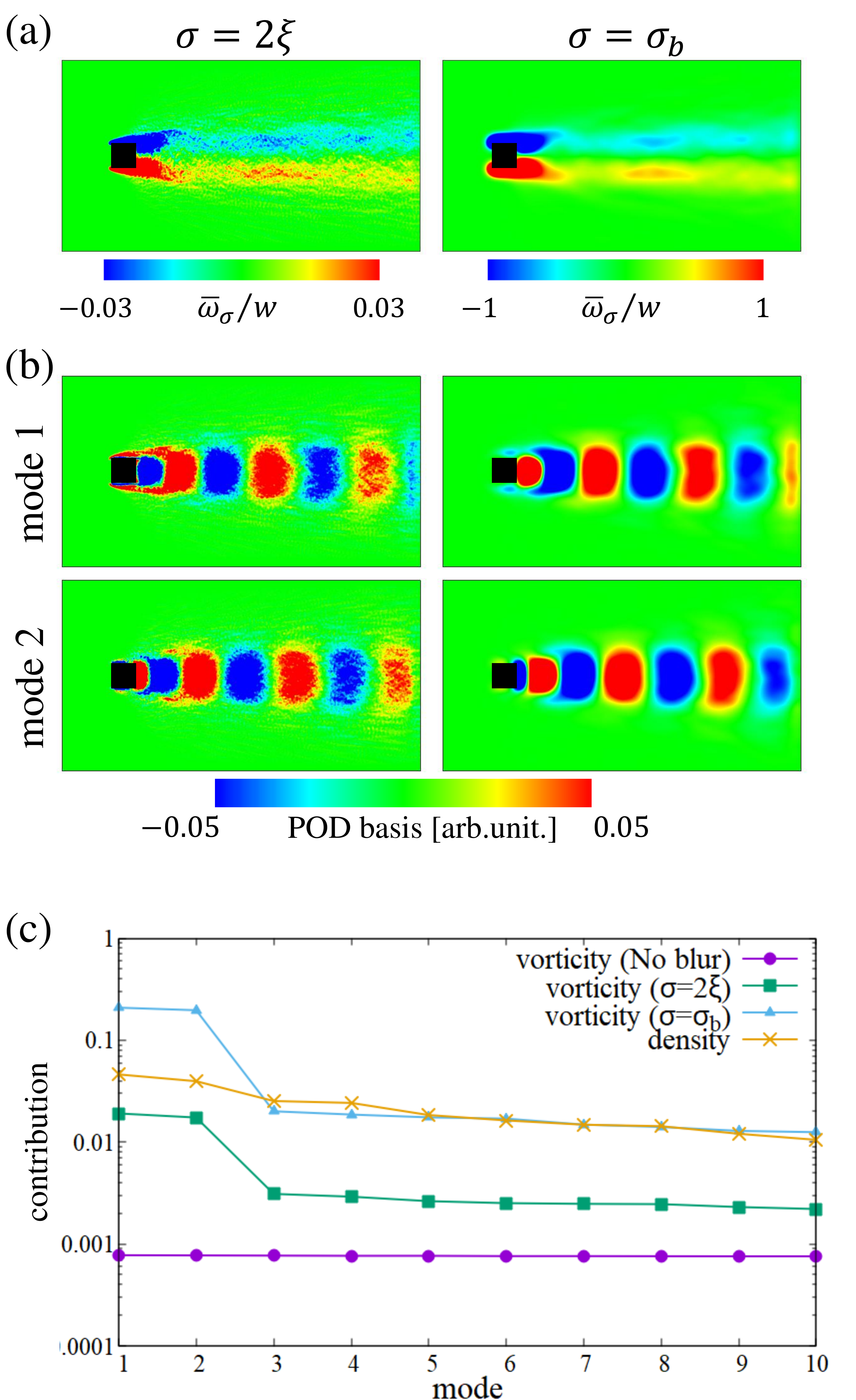}
\caption{
(Color online) (a, b) Results of filtered POD for the same flow ($Re_s=7.13$), as in Fig.~\ref{fig:POD}.
Left and right panels show plots by the Gaussian blur with deviation ranges of $2\xi$ and $\sigma_b$, repectively.
The mean filtered vorticity $\bar{\omega}_\sigma$ is rescaled by $w=\xi^2\kappa/2\pi \sigma^2 \Delta x^2$.
(c) 
Normalized eigenvalues (contribution) up to the 10th mode.
Filled marks show the results of unfiltered vorticity (circle) and filtered ones with deviation ranges $2\xi$ (square) and $\sigma_b$ (triangle).
The eigenvalues of the density POD demonstrated in Fig.~\ref{fig:POD} are plotted by cross marks for comparison.
Since the contributions of the two major modes are comparable, depending on the initial conditions or blurring method, mode 1 and mode 2 may be reversed.
}
\label{fig:filter}
\end{figure}

To address this issue, a coarse-graining filter, called a Gaussian blur, is applied to the vorticity distribution.
The filtered vorticity $\omega_\sigma$ with deviation range $\sigma$ is written as
\begin{equation}
    \omega_\sigma({\bm x})=\frac{1}{2\pi\sigma^2}\int d^2y \exp\left( -\frac{|\bm{y}|^2}{2\sigma^2} \right)\omega(\bm{x}-\bm{y})
    \label{eq:GFilter}
\end{equation}
This filter results in a broadened distribution of vorticity localized around a quantum vortex, but the total circulation is preserved.
Considering that density holes of radius $\sim\xi$ around quantum vortex axes led to the success of the density POD,
we first set $\sigma=2\xi$.
The results of the vorticity POD with this filtered vorticity $\omega_{2\xi}$ are shown in Fig.~\ref{fig:filter}.
The mean vorticity is antisymmetric and the major modes form a pair of symmetric patterns.
This patterns represent a periodic array of alternating vorticity with opposite sign circulation,
which is quite similar to the vorticity POD patterns of the classical KVS wake \cite{Vitkovicova2020POD,Dewanshu2022POD}.

One important problem with this method is that when typical vortex spacing becomes sufficiently larger than the healing length,
we have no clear pattern in the major modes as in the unfiltered POD.
We solve this problem as follows by determining the dispersion range $\sigma$ based on the idea of quasi-classical KVS illustrated in Fig.~\ref{fig:bundlewake}.

A clear indication of quasi-classical KVS must be visible in the major modes of the vorticity 
if each vortex bundle is coarse-grained as a single large vortex by applying appropriate filtering.
Namely, this could be achieved by taking $\sigma$ as the average spacing $\sigma_b$ between neighboring vortices in a vortex bundle:
\begin{equation}
    \sigma_b \equiv \sqrt{S_b/N_v}
    \label{eq:sigma_b}
\end{equation}
with the cross-sectional area $S_b$ of the bundle and the number $N_v$ of quantum vortices within.
As the magnitude of circulation of each vortex bundle that makes up the quasi-classical KVS is $\Gamma=\kappa N_v$, we have
\begin{equation}
    N_v= {\cal N} \equiv 2\sqrt{2}\frac{l}{d}\left( 1-\frac{l}{d}\mathit{St} \right) Re_s 
    \label{eq:Nv}
\end{equation}
with the bundle spacing $l$ in the flow direction.
Here, we used the result of the Kármán's theory \cite{Karman1912KVS},
$\Gamma=2\sqrt{2}lV$ with speed $V=U-fl=U-l\mathit{St}U/d$ of the bundle arrays,
and the superfluid Reynolds number
\begin{equation}
Re_s=dU/\kappa,
\end{equation}
assuming the circulation quantum $\kappa$ as an effective viscosity.

It is sufficient for our purposes if we could have a rough estimation for $\sigma_b$.
The Strouhal number in Eq.~\eqref{eq:Nv} is taken as $\mathit{St} = 0.15$, corresponding to the wake behind a square cylinder in classical fluids for high $Re$~\cite{Taha2015square}.
By assuming the spacing $b$ between the two bundle arrays [see Fig.~\ref{fig:bundlewake}] is of the same length as $d$, we substitute $d/l\sim b/l=\cosh^{-1}{(\sqrt{2})}/\pi \approx 0.2806$ and $S_b\sim d^2$, yielding
\begin{equation}
    \sigma_b/d = 0.3813/\sqrt{Re_s}.
    \label{eq:sigma_b/d}
\end{equation}
The POD analysis in terms of Eq.~\eqref{eq:GFilter} with $\sigma=\sigma_b$ captures the characteristics of KVS more clearly than previous results with $\sigma=2\xi$ [see right panels in Fig.~\ref{fig:filter}~(a,b)].
We obtained similar POD patterns in a wide range of $Re_s$.
What is noteworthy is that this filter significantly increases the contribution of the first and second modes compared to the subsequent modes [see FIG.~\ref{fig:filter}, (c)].
These facts demonstrate just that the quasi-classical KVS formed by vortex bundles is the dominant component of the flow, and clearly indicates the successful application of POD to the superfluid wake system.

\begin{figure}[thbp]
\centering
\includegraphics[width=8.5cm]{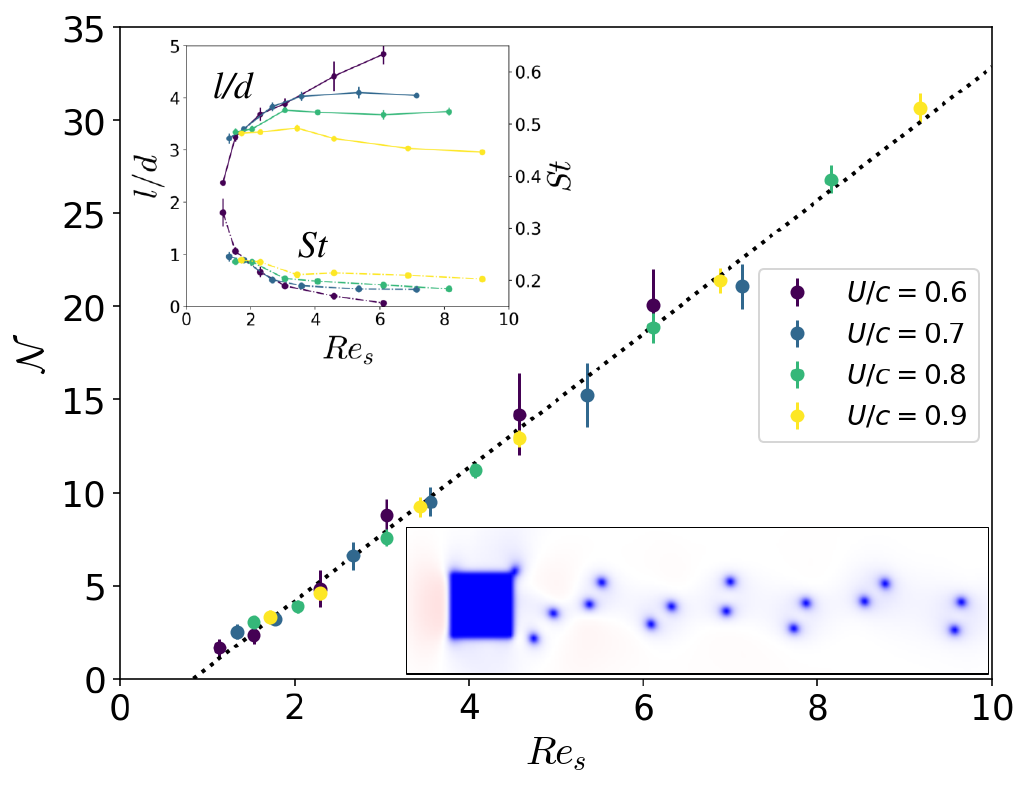}
\caption{
(Color online) Plots of ${\cal N}$ as a function of $Re_s$ for {\it irregular} wakes with different $d$ and $U$.
To compute Eq.~(\ref{eq:Nv}), plots of $l/d$ and $\mathit{St}$ in the inset are used
by taking ensemble averages of the spatial and temporal periods, $l$ and $2\pi/f$, respectively,
obtained from the first and second POD modes with $\sigma=\sigma_b$.
The error bars represent the standard deviation of the averages over 5 realizations of superfluid wakes with different initial conditions.
A dashed line shows the linear fitting function by the least-squares method,
suggesting that ${\cal N}$ is a universal function of $Re_s$; ${\cal N}=3.59 (Re_s-0.844)$.
A density snapshot of the numerical simulation with $Re_s=1.27$ ($d/\xi=16$ and $U/c=0.5$) is shown as an example of quantized KVS in lower right inset (not plotted).
}
\label{fig:Nv}
\end{figure}

\textit{Classification of superfluid wakes.}
Finally, we show that ${\cal N}$ defined by Eq.~\eqref{eq:Nv} can be an indicator classifying the \textit{irregular} superfluid wakes.
We numerically obtain $l/d$ and $\mathit{St}$ respectively from the spatial and temporal periods of the two most major modes with the dispersion range of Eq.~(\ref{eq:sigma_b/d}).
\footnote{(Supplemental Material) Details on the way to compute $l/d$ and $\mathit{St}$ can be accessed online.}
When we plotted ${\cal N}$ in this way over a wide range of parameters,
it is found by chance that it collapses to a single universal function of $Re_s$ as shown in Fig.~\ref{fig:Nv}.
Although $l/d$ and $\mathit{St}$ depend on $Re_s$ and $U/c$ [see the inset of Fig.~\ref{fig:Nv}],
substitution of them into Eq.~\eqref{eq:Nv} makes ${\cal N}$ dependent only on $Re_s$.
This suggests the existence of the Reynolds similitude described by the superfluid Reynolds number \cite{Takeuchi2024Rs,Christenhusz2024Rs}.
Fitting based on all the data agrees well with $\mathcal{N}=A(Re_s-B)$ with $A=3.59$ and $B=0.844$.
It should be noted that the fitting function is zero at $Re_s=B$.
Considering the physical meaning of $N_v$ in Eq.~(\ref{eq:Nv}),
$B$ suggests the critical superfluid Reynolds number,
below which no KVS are formed.
We could also find the {\it quantized} KVS composed of like-sign vortex pairs (lower right inset in Fig.~\ref{fig:Nv}) only for small sizes {\it e.g.}, $d/\xi=12, 16$ with $Re_s\sim 1$.
Then we have $\mathcal{N}\sim 1$,
consistently suggesting that each vortex bundle holds a few vortices.

It is instructive to qualitatively compare our result to the previous work.
According to the phase diagram of Fig. 3 in Ref.~\cite{Sasaki2010wake},
they obtained the {\it quantized} KVS consisting of like-sign vortex pairs for $Re_s\sim 0.1$.
Substitution of this into our fitting function results in negative ${\cal N}$.
This unphysical result suggests that the universal function should be determined depending on shape of the obstacle considered.
We expect the quasi-classical KVS to be well defined when $Re_s$ ({\it i.e.}, ${\cal N}$) is sufficiently large.
Therefore, the fact that the fitting function does not pass through the origin can be attributed to a kind of correction due to quantum effects for small $Re_s$ \cite{Takeuchi2024Rs}.

\textit{Summary and discussion.}
Our POD analysis succeeded to find the major modes of quasi-classical KVS in superfluid turbulent wakes,
considered \textit{irregular} in previous studies \cite{Sasaki2010wake,Reeves2015Rs}.
Our technique of filtered vorticity can be extended to superfluid wakes behind obstacles with different shapes in two and three dimensions.
The density POD can be useful for experimental data analyses
since density is observable in ultra-cold atomic gases.
It is heuristically found that the dimensionless number ${\cal N}$ defined by Eq.~\eqref{eq:Nv} behaves as a universal function that depends only on the superfluid Reynolds number $Re_s$ in the parameter regime ($1\lesssim Re_s \lesssim  10$) we have studied.
To investigate the universal behavior of quasi-classical KVS for $Re_s > 10$ is our interesting future work
as KVS is realized typically for $Re>10$ in classical fluids.

\begin{acknowledgments}
We thank M. Nakata, S. Sato, Y. Ohmichi, T. Tanogami, and S. Goto for useful discussions and information.
This research was conducted using the FUJITSU Supercomputer PRIMEHPC FX1000 and FUJITSU Server PRIMERGY GX2570 (Wisteria/BDEC-01) at the Information Technology Center, The University of Tokyo.
This work is supported by JSPS KAKENHI Grants Nos. JP18KK0391 and JP20H01842; and JST, PRESTO (Japan) Grant No. JPMJPR23O5.
\end{acknowledgments}

\bibliography{cite}

%apsrev4-2.bst 2019-01-14 (MD) hand-edited version of apsrev4-1.bst
%Control: key (0)
%Control: author (8) initials jnrlst
%Control: editor formatted (1) identically to author
%Control: production of article title (0) allowed
%Control: page (0) single
%Control: year (1) truncated
%Control: production of eprint (0) enabled
\begin{thebibliography}{41}%
\makeatletter
\providecommand \@ifxundefined [1]{%
 \@ifx{#1\undefined}
}%
\providecommand \@ifnum [1]{%
 \ifnum #1\expandafter \@firstoftwo
 \else \expandafter \@secondoftwo
 \fi
}%
\providecommand \@ifx [1]{%
 \ifx #1\expandafter \@firstoftwo
 \else \expandafter \@secondoftwo
 \fi
}%
\providecommand \natexlab [1]{#1}%
\providecommand \enquote  [1]{``#1''}%
\providecommand \bibnamefont  [1]{#1}%
\providecommand \bibfnamefont [1]{#1}%
\providecommand \citenamefont [1]{#1}%
\providecommand \href@noop [0]{\@secondoftwo}%
\providecommand \href [0]{\begingroup \@sanitize@url \@href}%
\providecommand \@href[1]{\@@startlink{#1}\@@href}%
\providecommand \@@href[1]{\endgroup#1\@@endlink}%
\providecommand \@sanitize@url [0]{\catcode `\\12\catcode `\$12\catcode `\&12\catcode `\#12\catcode `\^12\catcode `\_12\catcode `\%12\relax}%
\providecommand \@@startlink[1]{}%
\providecommand \@@endlink[0]{}%
\providecommand \url  [0]{\begingroup\@sanitize@url \@url }%
\providecommand \@url [1]{\endgroup\@href {#1}{\urlprefix }}%
\providecommand \urlprefix  [0]{URL }%
\providecommand \Eprint [0]{\href }%
\providecommand \doibase [0]{https://doi.org/}%
\providecommand \selectlanguage [0]{\@gobble}%
\providecommand \bibinfo  [0]{\@secondoftwo}%
\providecommand \bibfield  [0]{\@secondoftwo}%
\providecommand \translation [1]{[#1]}%
\providecommand \BibitemOpen [0]{}%
\providecommand \bibitemStop [0]{}%
\providecommand \bibitemNoStop [0]{.\EOS\space}%
\providecommand \EOS [0]{\spacefactor3000\relax}%
\providecommand \BibitemShut  [1]{\csname bibitem#1\endcsname}%
\let\auto@bib@innerbib\@empty
%</preamble>
\bibitem [{\citenamefont {Chapin}\ and\ \citenamefont {Nicolelis}(1999)}]{Neuroscience1999}%
  \BibitemOpen
  \bibfield  {author} {\bibinfo {author} {\bibfnamefont {J.~K.}\ \bibnamefont {Chapin}}\ and\ \bibinfo {author} {\bibfnamefont {M.~A.}\ \bibnamefont {Nicolelis}},\ }\bibfield  {title} {\bibinfo {title} {{Principal component analysis of neuronal ensemble activity reveals multidimensional somatosensory representations}},\ }\href {https://doi.org/10.1016/S0165-0270(99)00130-2} {\bibfield  {journal} {\bibinfo  {journal} {J. Neurosci. Methods}\ }\textbf {\bibinfo {volume} {94}},\ \bibinfo {pages} {121} (\bibinfo {year} {1999})}\BibitemShut {NoStop}%
\bibitem [{\citenamefont {Hannachi}\ \emph {et~al.}(2006)\citenamefont {Hannachi}, \citenamefont {Jolliffe}, \citenamefont {Stephenson},\ and\ \citenamefont {Trendafilov}}]{Atmospheric2006}%
  \BibitemOpen
  \bibfield  {author} {\bibinfo {author} {\bibfnamefont {A.}~\bibnamefont {Hannachi}}, \bibinfo {author} {\bibfnamefont {I.~T.}\ \bibnamefont {Jolliffe}}, \bibinfo {author} {\bibfnamefont {D.~B.}\ \bibnamefont {Stephenson}},\ and\ \bibinfo {author} {\bibfnamefont {N.}~\bibnamefont {Trendafilov}},\ }\bibfield  {title} {\bibinfo {title} {{In search of simple structures in climate: simplifying EOFs}},\ }\href {https://doi.org/10.1002/joc.1243} {\bibfield  {journal} {\bibinfo  {journal} {Int. J. Clim.}\ }\textbf {\bibinfo {volume} {26}},\ \bibinfo {pages} {7} (\bibinfo {year} {2006})}\BibitemShut {NoStop}%
\bibitem [{\citenamefont {Gill}\ \emph {et~al.}(2014)\citenamefont {Gill}, \citenamefont {Purnell}, \citenamefont {Crumpton}, \citenamefont {Brown}, \citenamefont {Gostling}, \citenamefont {Stampanoni},\ and\ \citenamefont {Rayfield}}]{paleontology2014}%
  \BibitemOpen
  \bibfield  {author} {\bibinfo {author} {\bibfnamefont {P.~G.}\ \bibnamefont {Gill}}, \bibinfo {author} {\bibfnamefont {M.~A.}\ \bibnamefont {Purnell}}, \bibinfo {author} {\bibfnamefont {N.}~\bibnamefont {Crumpton}}, \bibinfo {author} {\bibfnamefont {K.~R.}\ \bibnamefont {Brown}}, \bibinfo {author} {\bibfnamefont {N.~J.}\ \bibnamefont {Gostling}}, \bibinfo {author} {\bibfnamefont {M.}~\bibnamefont {Stampanoni}},\ and\ \bibinfo {author} {\bibfnamefont {E.~J.}\ \bibnamefont {Rayfield}},\ }\bibfield  {title} {\bibinfo {title} {{Dietary specializations and diversity in feeding ecology of the earliest stem mammals}},\ }\href {https://doi.org/10.1038/nature13622} {\bibfield  {journal} {\bibinfo  {journal} {Nature}\ }\textbf {\bibinfo {volume} {512}},\ \bibinfo {pages} {303} (\bibinfo {year} {2014})}\BibitemShut {NoStop}%
\bibitem [{\citenamefont {Lumley}(1967)}]{Lumley1967POD}%
  \BibitemOpen
  \bibfield  {author} {\bibinfo {author} {\bibfnamefont {J.~L.}\ \bibnamefont {Lumley}},\ }\bibfield  {title} {\bibinfo {title} {{The structure of inhomogeneous turbulence.}},\ }\href {https://scholar.google.com/scholar_lookup?title=The%20structure%20of%20inhomogeneous%20turbulent%20flows&author=J.L.%20Lumley&publication_year=1967} {\bibfield  {journal} {\bibinfo  {journal} {Atmospheric Turbulence and Radio Wave propagation}\ ,\ \bibinfo {pages} {166}} (\bibinfo {year} {1967})}\BibitemShut {NoStop}%
\bibitem [{\citenamefont {Rehimi}\ \emph {et~al.}(2008)\citenamefont {Rehimi}, \citenamefont {Aloui}, \citenamefont {Ben~Nasrallah}, \citenamefont {Doubliez},\ and\ \citenamefont {Legrand}}]{Rehimi2008POD}%
  \BibitemOpen
  \bibfield  {author} {\bibinfo {author} {\bibfnamefont {F.}~\bibnamefont {Rehimi}}, \bibinfo {author} {\bibfnamefont {F.}~\bibnamefont {Aloui}}, \bibinfo {author} {\bibfnamefont {S.}~\bibnamefont {Ben~Nasrallah}}, \bibinfo {author} {\bibfnamefont {L.}~\bibnamefont {Doubliez}},\ and\ \bibinfo {author} {\bibfnamefont {J.}~\bibnamefont {Legrand}},\ }\bibfield  {title} {\bibinfo {title} {{Experimental investigation of a confined flow downstream of a circular cylinder centred between two parallel walls}},\ }\href {https://doi.org/10.1016/j.jfluidstructs.2007.12.011} {\bibfield  {journal} {\bibinfo  {journal} {J. Fluids Struct.}\ }\textbf {\bibinfo {volume} {24}},\ \bibinfo {pages} {855} (\bibinfo {year} {2008})}\BibitemShut {NoStop}%
\bibitem [{\citenamefont {Wang}\ \emph {et~al.}(2014)\citenamefont {Wang}, \citenamefont {Cao},\ and\ \citenamefont {Zhow}}]{Wang2014POD}%
  \BibitemOpen
  \bibfield  {author} {\bibinfo {author} {\bibfnamefont {H.~F.}\ \bibnamefont {Wang}}, \bibinfo {author} {\bibfnamefont {L.}~\bibnamefont {Cao}, \bibfnamefont {H}},\ and\ \bibinfo {author} {\bibfnamefont {Y.}~\bibnamefont {Zhow}},\ }\bibfield  {title} {\bibinfo {title} {{POD analysis of a finite-length cylinder near wake}},\ }\href {https://doi.org/10.1007/s00348-014-1790-9} {\bibfield  {journal} {\bibinfo  {journal} {Exp. Fluids}\ }\textbf {\bibinfo {volume} {55}} (\bibinfo {year} {2014})}\BibitemShut {NoStop}%
\bibitem [{\citenamefont {Samani}\ and\ \citenamefont {Bergstrom}(2015)}]{Samani2015POD}%
  \BibitemOpen
  \bibfield  {author} {\bibinfo {author} {\bibfnamefont {S.}~\bibnamefont {Samani}}\ and\ \bibinfo {author} {\bibfnamefont {D.~J.}\ \bibnamefont {Bergstrom}},\ }\bibfield  {title} {\bibinfo {title} {{Effect of a wall on the wake dynamics of an infinite square cylinder}},\ }\href {https://doi.org/10.1016/j.ijheatfluidflow.2015.07.016} {\bibfield  {journal} {\bibinfo  {journal} {Int. J. Heat Fluid Flow}\ }\textbf {\bibinfo {volume} {55}},\ \bibinfo {pages} {158} (\bibinfo {year} {2015})}\BibitemShut {NoStop}%
\bibitem [{\citenamefont {Reynolds}(1883)}]{Reynolds1883}%
  \BibitemOpen
  \bibfield  {author} {\bibinfo {author} {\bibfnamefont {O.}~\bibnamefont {Reynolds}},\ }\bibfield  {title} {\bibinfo {title} {{XXIX. An experimental investigation of the circumstances which determine whether the motion of water shall be direct or sinuous, and of the law of resistance in parallel channels}},\ }\href {https://doi.org/10.1098/rstl.1883.0029} {\bibfield  {journal} {\bibinfo  {journal} {Philos.Trans.R.Soc.London}\ }\textbf {\bibinfo {volume} {174}},\ \bibinfo {pages} {935} (\bibinfo {year} {1883})}\BibitemShut {NoStop}%
\bibitem [{\citenamefont {Williamson}(1996)}]{Williamson1996KVS}%
  \BibitemOpen
  \bibfield  {author} {\bibinfo {author} {\bibfnamefont {C.~H.~K.}\ \bibnamefont {Williamson}},\ }\bibfield  {title} {\bibinfo {title} {{Vortex Dynamics in the Cylinder Wake}},\ }\href {https://doi.org/10.1146/annurev.fl.28.010196.002401} {\bibfield  {journal} {\bibinfo  {journal} {Annu. Rev. Fluid Mech.}\ }\textbf {\bibinfo {volume} {28}},\ \bibinfo {pages} {477} (\bibinfo {year} {1996})}\BibitemShut {NoStop}%
\bibitem [{\citenamefont {Vitkovicova}\ \emph {et~al.}(2020)\citenamefont {Vitkovicova}, \citenamefont {Yokoi},\ and\ \citenamefont {Hyhlik}}]{Vitkovicova2020POD}%
  \BibitemOpen
  \bibfield  {author} {\bibinfo {author} {\bibfnamefont {R.}~\bibnamefont {Vitkovicova}}, \bibinfo {author} {\bibfnamefont {Y.}~\bibnamefont {Yokoi}},\ and\ \bibinfo {author} {\bibfnamefont {T.}~\bibnamefont {Hyhlik}},\ }\bibfield  {title} {\bibinfo {title} {{Identification of structures and mechanisms in a flow field by POD analysis for input data obtained from visualization and PIV}},\ }\href {https://doi.org/10.1007/s00348-020-03005-6} {\bibfield  {journal} {\bibinfo  {journal} {Exp. Fluids}\ }\textbf {\bibinfo {volume} {61}} (\bibinfo {year} {2020})}\BibitemShut {NoStop}%
\bibitem [{\citenamefont {Deep}\ \emph {et~al.}(2022)\citenamefont {Deep}, \citenamefont {Sahasranaman},\ and\ \citenamefont {Senthilkumar}}]{Dewanshu2022POD}%
  \BibitemOpen
  \bibfield  {author} {\bibinfo {author} {\bibfnamefont {D.}~\bibnamefont {Deep}}, \bibinfo {author} {\bibfnamefont {A.}~\bibnamefont {Sahasranaman}},\ and\ \bibinfo {author} {\bibfnamefont {S.}~\bibnamefont {Senthilkumar}},\ }\bibfield  {title} {\bibinfo {title} {{POD analysis of the wake behind a circular cylinder with splitter plate}},\ }\href {https://doi.org/10.1016/j.euromechflu.2021.12.010} {\bibfield  {journal} {\bibinfo  {journal} {Eur. J. Mech. B Fluids.}\ }\textbf {\bibinfo {volume} {93}},\ \bibinfo {pages} {1} (\bibinfo {year} {2022})}\BibitemShut {NoStop}%
\bibitem [{\citenamefont {Sasaki}\ \emph {et~al.}(2010)\citenamefont {Sasaki}, \citenamefont {Suzuki},\ and\ \citenamefont {Saito}}]{Sasaki2010wake}%
  \BibitemOpen
  \bibfield  {author} {\bibinfo {author} {\bibfnamefont {K.}~\bibnamefont {Sasaki}}, \bibinfo {author} {\bibfnamefont {N.}~\bibnamefont {Suzuki}},\ and\ \bibinfo {author} {\bibfnamefont {H.}~\bibnamefont {Saito}},\ }\bibfield  {title} {\bibinfo {title} {{Bénard–von Kármán Vortex Street in a Bose-Einstein Condensate}},\ }\href {https://doi.org/10.1103/PhysRevLett.104.150404} {\bibfield  {journal} {\bibinfo  {journal} {Phys. Rev. Lett.}\ }\textbf {\bibinfo {volume} {104}},\ \bibinfo {pages} {15404} (\bibinfo {year} {2010})}\BibitemShut {NoStop}%
\bibitem [{\citenamefont {Kwon}\ \emph {et~al.}(2016)\citenamefont {Kwon}, \citenamefont {Kim}, \citenamefont {Seo},\ and\ \citenamefont {Shin}}]{Kwon2016QKarman}%
  \BibitemOpen
  \bibfield  {author} {\bibinfo {author} {\bibfnamefont {W.~J.}\ \bibnamefont {Kwon}}, \bibinfo {author} {\bibfnamefont {J.~H.}\ \bibnamefont {Kim}}, \bibinfo {author} {\bibfnamefont {S.~W.}\ \bibnamefont {Seo}},\ and\ \bibinfo {author} {\bibfnamefont {Y.}~\bibnamefont {Shin}},\ }\bibfield  {title} {\bibinfo {title} {{Observation of von Kármán Vortex Street in an Atomic Superfluid Gas}},\ }\href {https://doi.org/10.1103/PhysRevLett.117.245301} {\bibfield  {journal} {\bibinfo  {journal} {Phys. Rev. Lett.}\ }\textbf {\bibinfo {volume} {117}},\ \bibinfo {pages} {245301} (\bibinfo {year} {2016})}\BibitemShut {NoStop}%
\bibitem [{\citenamefont {Nore}\ \emph {et~al.}(2000{\natexlab{a}})\citenamefont {Nore}, \citenamefont {Abid},\ and\ \citenamefont {Brachet}}]{Nore1997bundle}%
  \BibitemOpen
  \bibfield  {author} {\bibinfo {author} {\bibfnamefont {C.}~\bibnamefont {Nore}}, \bibinfo {author} {\bibfnamefont {M.}~\bibnamefont {Abid}},\ and\ \bibinfo {author} {\bibfnamefont {M.~E.}\ \bibnamefont {Brachet}},\ }\bibfield  {title} {\bibinfo {title} {{Kolmogorov Turbulence in Low-Temperature Superflows}},\ }\href {https://doi.org/10.1103/PhysRevLett.78.3896} {\bibfield  {journal} {\bibinfo  {journal} {Phys. Rev. Lett.}\ }\textbf {\bibinfo {volume} {78}},\ \bibinfo {pages} {3896} (\bibinfo {year} {2000}{\natexlab{a}})}\BibitemShut {NoStop}%
\bibitem [{\citenamefont {Nore}\ \emph {et~al.}(2000{\natexlab{b}})\citenamefont {Nore}, \citenamefont {Huepe},\ and\ \citenamefont {Brachet}}]{Nore2000Rs}%
  \BibitemOpen
  \bibfield  {author} {\bibinfo {author} {\bibfnamefont {C.}~\bibnamefont {Nore}}, \bibinfo {author} {\bibfnamefont {C.}~\bibnamefont {Huepe}},\ and\ \bibinfo {author} {\bibfnamefont {M.~E.}\ \bibnamefont {Brachet}},\ }\bibfield  {title} {\bibinfo {title} {{Subcritical Dissipation in Three-Dimensional Superflows}},\ }\href {https://doi.org/10.1103/PhysRevLett.84.2191} {\bibfield  {journal} {\bibinfo  {journal} {Phys. Rev. Lett.}\ }\textbf {\bibinfo {volume} {84}},\ \bibinfo {pages} {2191} (\bibinfo {year} {2000}{\natexlab{b}})}\BibitemShut {NoStop}%
\bibitem [{\citenamefont {Volovik}(2003)}]{Volovik2003}%
  \BibitemOpen
  \bibfield  {author} {\bibinfo {author} {\bibfnamefont {G.~E.}\ \bibnamefont {Volovik}},\ }\bibfield  {title} {\bibinfo {title} {{Classical and Quantum Regimes of Superfluid Turbulence}},\ }\href {https://doi.org/10.1134/1.1641478} {\bibfield  {journal} {\bibinfo  {journal} {JETP Letters}\ }\textbf {\bibinfo {volume} {78}},\ \bibinfo {pages} {533} (\bibinfo {year} {2003})}\BibitemShut {NoStop}%
\bibitem [{\citenamefont {Reeves}\ \emph {et~al.}(2015)\citenamefont {Reeves}, \citenamefont {Billam}, \citenamefont {Anderson},\ and\ \citenamefont {Bradley}}]{Reeves2015Rs}%
  \BibitemOpen
  \bibfield  {author} {\bibinfo {author} {\bibfnamefont {M.~T.}\ \bibnamefont {Reeves}}, \bibinfo {author} {\bibfnamefont {T.~P.}\ \bibnamefont {Billam}}, \bibinfo {author} {\bibfnamefont {B.~P.}\ \bibnamefont {Anderson}},\ and\ \bibinfo {author} {\bibfnamefont {A.~S.}\ \bibnamefont {Bradley}},\ }\bibfield  {title} {\bibinfo {title} {{Identifying a Superfluid Reynolds Number via Dynamical Similarity}},\ }\href {https://doi.org/10.1103/PhysRevLett.114.155302} {\bibfield  {journal} {\bibinfo  {journal} {Phys. Rev. Lett.}\ }\textbf {\bibinfo {volume} {114}},\ \bibinfo {pages} {155302} (\bibinfo {year} {2015})}\BibitemShut {NoStop}%
\bibitem [{\citenamefont {Schoepe}(2015)}]{Schoepe2015Res}%
  \BibitemOpen
  \bibfield  {author} {\bibinfo {author} {\bibfnamefont {W.}~\bibnamefont {Schoepe}},\ }\bibfield  {title} {\bibinfo {title} {{Superfluid Reynolds Number and the Transition from Potential Flow to Turbulence in Superfluid 4He at Millikelvin Temperatures}},\ }\href {https://doi.org/10.1134/S0021364015140106} {\bibfield  {journal} {\bibinfo  {journal} {JETP Letters}\ }\textbf {\bibinfo {volume} {102}},\ \bibinfo {pages} {105} (\bibinfo {year} {2015})}\BibitemShut {NoStop}%
\bibitem [{\citenamefont {Lim}\ \emph {et~al.}(2022)\citenamefont {Lim}, \citenamefont {Lee}, \citenamefont {Goo}, \citenamefont {Bae},\ and\ \citenamefont {Shin}}]{Lim2022Res}%
  \BibitemOpen
  \bibfield  {author} {\bibinfo {author} {\bibfnamefont {Y.}~\bibnamefont {Lim}}, \bibinfo {author} {\bibfnamefont {Y.}~\bibnamefont {Lee}}, \bibinfo {author} {\bibfnamefont {J.}~\bibnamefont {Goo}}, \bibinfo {author} {\bibfnamefont {D.}~\bibnamefont {Bae}},\ and\ \bibinfo {author} {\bibfnamefont {Y.}~\bibnamefont {Shin}},\ }\bibfield  {title} {\bibinfo {title} {{Vortex shedding frequency of a moving obstacle in a Bose–Einstein condensate}},\ }\href {https://doi.org/10.1088/1367-2630/ac8656} {\bibfield  {journal} {\bibinfo  {journal} {New J. Phys.}\ }\textbf {\bibinfo {volume} {24}},\ \bibinfo {pages} {083020} (\bibinfo {year} {2022})}\BibitemShut {NoStop}%
\bibitem [{\citenamefont {Schoepe}(2022)}]{Schoepe2022Res}%
  \BibitemOpen
  \bibfield  {author} {\bibinfo {author} {\bibfnamefont {W.}~\bibnamefont {Schoepe}},\ }\bibfield  {title} {\bibinfo {title} {{Vortex Shedding from a Microsphere Oscillating in Superfluid 4He at mK Temperatures and from a Laser Beam Moving in a Bose–Einstein Condensate}},\ }\href {https://doi.org/10.1007/s10909-022-02716-w} {\bibfield  {journal} {\bibinfo  {journal} {J. Low Temp. Phys.}\ }\textbf {\bibinfo {volume} {210}},\ \bibinfo {pages} {539} (\bibinfo {year} {2022})}\BibitemShut {NoStop}%
\bibitem [{\citenamefont {Wang}\ \emph {et~al.}(2022)\citenamefont {Wang}, \citenamefont {Li}, \citenamefont {Ren}, \citenamefont {Fan}, \citenamefont {Zhou}, \citenamefont {Meng}, \citenamefont {Wan}, \citenamefont {Zhang}, \citenamefont {Shao},\ and\ \citenamefont {Shi}}]{Wang2022QKVS}%
  \BibitemOpen
  \bibfield  {author} {\bibinfo {author} {\bibfnamefont {J.}~\bibnamefont {Wang}}, \bibinfo {author} {\bibfnamefont {X.}~\bibnamefont {Li}}, \bibinfo {author} {\bibfnamefont {X.}~\bibnamefont {Ren}}, \bibinfo {author} {\bibfnamefont {X.}~\bibnamefont {Fan}}, \bibinfo {author} {\bibfnamefont {Y.}~\bibnamefont {Zhou}}, \bibinfo {author} {\bibfnamefont {H.}~\bibnamefont {Meng}}, \bibinfo {author} {\bibfnamefont {X.}~\bibnamefont {Wan}}, \bibinfo {author} {\bibfnamefont {J.}~\bibnamefont {Zhang}}, \bibinfo {author} {\bibfnamefont {K.}~\bibnamefont {Shao}},\ and\ \bibinfo {author} {\bibfnamefont {Y.}~\bibnamefont {Shi}},\ }\bibfield  {title} {\bibinfo {title} {{Quantum Kármán vortex street in an immiscible two-component Bose–Einstein condensate}},\ }\href {https://doi.org/10.1140/epjp/s13360-022-03420-0} {\bibfield  {journal} {\bibinfo  {journal} {Eur. Phys. J. Plus}\ }\textbf {\bibinfo {volume} {137}} (\bibinfo {year} {2022})}\BibitemShut {NoStop}%
\bibitem [{\citenamefont {Skrbek}\ \emph {et~al.}(2024)\citenamefont {Skrbek}, \citenamefont {Schmoranzer},\ and\ \citenamefont {Sreenivasan}}]{Skrbek2024Res}%
  \BibitemOpen
  \bibfield  {author} {\bibinfo {author} {\bibfnamefont {L.}~\bibnamefont {Skrbek}}, \bibinfo {author} {\bibfnamefont {D.}~\bibnamefont {Schmoranzer}},\ and\ \bibinfo {author} {\bibfnamefont {K.~R.}\ \bibnamefont {Sreenivasan}},\ }\bibfield  {title} {\bibinfo {title} {{Phenomenology of transition to quantum turbulence in flows of superfluid helium}},\ }\href {https://doi.org/10.1073/pnas.2302256121} {\bibfield  {journal} {\bibinfo  {journal} {PNAS}\ }\textbf {\bibinfo {volume} {121}},\ \bibinfo {pages} {e2302256121} (\bibinfo {year} {2024})}\BibitemShut {NoStop}%
\bibitem [{\citenamefont {Takeuchi}(2024)}]{Takeuchi2024Rs}%
  \BibitemOpen
  \bibfield  {author} {\bibinfo {author} {\bibfnamefont {H.}~\bibnamefont {Takeuchi}},\ }\bibfield  {title} {\bibinfo {title} {{Quantum viscosity and the Reynolds similitude of a pure superfluid}},\ }\href {https://doi.org/10.1103/PhysRevB.109.L020502} {\bibfield  {journal} {\bibinfo  {journal} {Phys. Rev. B}\ }\textbf {\bibinfo {volume} {109}},\ \bibinfo {pages} {L020502} (\bibinfo {year} {2024})}\BibitemShut {NoStop}%
\bibitem [{\citenamefont {Vinen}\ and\ \citenamefont {Niemela}(2002)}]{Vinen2002}%
  \BibitemOpen
  \bibfield  {author} {\bibinfo {author} {\bibfnamefont {W.~F.}\ \bibnamefont {Vinen}}\ and\ \bibinfo {author} {\bibfnamefont {J.~J.}\ \bibnamefont {Niemela}},\ }\bibfield  {title} {\bibinfo {title} {{Quantum Turbulence}},\ }\href {https://doi.org/10.1023/A:1019695418590} {\bibfield  {journal} {\bibinfo  {journal} {J. Low Temp. Phys.}\ }\textbf {\bibinfo {volume} {128}},\ \bibinfo {pages} {167} (\bibinfo {year} {2002})}\BibitemShut {NoStop}%
\bibitem [{\citenamefont {Batchelor}(1967)}]{Batchelor1967}%
  \BibitemOpen
  \bibfield  {author} {\bibinfo {author} {\bibfnamefont {G.~K.}\ \bibnamefont {Batchelor}},\ }\href@noop {} {\emph {\bibinfo {title} {{An introduction to fluid dynamics}}}}\ (\bibinfo  {publisher} {Cambridge university press},\ \bibinfo {year} {1967})\BibitemShut {NoStop}%
\bibitem [{\citenamefont {Kwon}\ \emph {et~al.}(2015)\citenamefont {Kwon}, \citenamefont {Moon}, \citenamefont {Seo},\ and\ \citenamefont {Shin}}]{Kwon2015Uc}%
  \BibitemOpen
  \bibfield  {author} {\bibinfo {author} {\bibfnamefont {W.~J.}\ \bibnamefont {Kwon}}, \bibinfo {author} {\bibfnamefont {G.}~\bibnamefont {Moon}}, \bibinfo {author} {\bibfnamefont {S.~W.}\ \bibnamefont {Seo}},\ and\ \bibinfo {author} {\bibfnamefont {Y.}~\bibnamefont {Shin}},\ }\bibfield  {title} {\bibinfo {title} {{Critical velocity for vortex shedding in a Bose-Einstein condensate}},\ }\href {https://doi.org/10.1103/PhysRevA.91.053615} {\bibfield  {journal} {\bibinfo  {journal} {Phys. Rev. A}\ }\textbf {\bibinfo {volume} {91}},\ \bibinfo {pages} {053615} (\bibinfo {year} {2015})}\BibitemShut {NoStop}%
\bibitem [{\citenamefont {Rica}(2001)}]{Rica2001Uc}%
  \BibitemOpen
  \bibfield  {author} {\bibinfo {author} {\bibfnamefont {S.}~\bibnamefont {Rica}},\ }\bibfield  {title} {\bibinfo {title} {{A remark on the critical speed for vortex nucleation in the nonlinear Schrödinger equation}},\ }\href {https://doi.org/10.1016/S0167-2789(00)00168-8} {\bibfield  {journal} {\bibinfo  {journal} {Physica D}\ }\textbf {\bibinfo {volume} {148}},\ \bibinfo {pages} {221} (\bibinfo {year} {2001})}\BibitemShut {NoStop}%
\bibitem [{\citenamefont {Kwak}\ \emph {et~al.}(2023)\citenamefont {Kwak}, \citenamefont {Jung},\ and\ \citenamefont {Y.}}]{Kwak2023Uc}%
  \BibitemOpen
  \bibfield  {author} {\bibinfo {author} {\bibfnamefont {H.}~\bibnamefont {Kwak}}, \bibinfo {author} {\bibfnamefont {J.~H.}\ \bibnamefont {Jung}},\ and\ \bibinfo {author} {\bibfnamefont {S.}~\bibnamefont {Y.}},\ }\bibfield  {title} {\bibinfo {title} {{Minimum critical velocity of a Gaussian obstacle in a Bose-Einstein condensate}},\ }\href {https://doi.org/10.1103/PhysRevA.107.023310} {\bibfield  {journal} {\bibinfo  {journal} {Phys. Rev. A}\ }\textbf {\bibinfo {volume} {107}},\ \bibinfo {pages} {023310} (\bibinfo {year} {2023})}\BibitemShut {NoStop}%
\bibitem [{\citenamefont {Frisch}\ \emph {et~al.}(1992)\citenamefont {Frisch}, \citenamefont {Pomeau},\ and\ \citenamefont {Rica}}]{Frisch1992wake}%
  \BibitemOpen
  \bibfield  {author} {\bibinfo {author} {\bibfnamefont {T.}~\bibnamefont {Frisch}}, \bibinfo {author} {\bibfnamefont {Y.}~\bibnamefont {Pomeau}},\ and\ \bibinfo {author} {\bibfnamefont {S.}~\bibnamefont {Rica}},\ }\bibfield  {title} {\bibinfo {title} {{Transition to dissipation in a model of superflow}},\ }\href {https://doi.org/10.1103/PhysRevLett.69.1644} {\bibfield  {journal} {\bibinfo  {journal} {Phys. Rev. Lett.}\ }\textbf {\bibinfo {volume} {69}},\ \bibinfo {pages} {1644} (\bibinfo {year} {1992})}\BibitemShut {NoStop}%
\bibitem [{\citenamefont {Pham}\ \emph {et~al.}(2005)\citenamefont {Pham}, \citenamefont {Nore},\ and\ \citenamefont {Étienne Brachet}}]{Pham2005Uc}%
  \BibitemOpen
  \bibfield  {author} {\bibinfo {author} {\bibfnamefont {C.-T.}\ \bibnamefont {Pham}}, \bibinfo {author} {\bibfnamefont {C.}~\bibnamefont {Nore}},\ and\ \bibinfo {author} {\bibfnamefont {M.}~\bibnamefont {Étienne Brachet}},\ }\bibfield  {title} {\bibinfo {title} {{Boundary layers and emitted excitations in nonlinear Schrödinger superflow past a disk}},\ }\href {https://doi.org/10.1016/j.physd.2005.07.013} {\bibfield  {journal} {\bibinfo  {journal} {Physica D}\ }\textbf {\bibinfo {volume} {210}},\ \bibinfo {pages} {203} (\bibinfo {year} {2005})}\BibitemShut {NoStop}%
\bibitem [{\citenamefont {Kokubo}\ \emph {et~al.}(2025)\citenamefont {Kokubo}, \citenamefont {Takeuchi},\ and\ \citenamefont {Kasamatsu}}]{Kokubo2025Uc}%
  \BibitemOpen
  \bibfield  {author} {\bibinfo {author} {\bibfnamefont {H.}~\bibnamefont {Kokubo}}, \bibinfo {author} {\bibfnamefont {H.}~\bibnamefont {Takeuchi}},\ and\ \bibinfo {author} {\bibfnamefont {K.}~\bibnamefont {Kasamatsu}},\ }\bibfield  {title} {\bibinfo {title} {{Critical velocity for wake vortex generation behind a plate in a superflow}},\ }\href {https://doi.org/10.1103/PhysRevA.111.043314} {\bibfield  {journal} {\bibinfo  {journal} {Phys. Rev. A}\ }\textbf {\bibinfo {volume} {111}},\ \bibinfo {pages} {043314} (\bibinfo {year} {2025})}\BibitemShut {NoStop}%
\bibitem [{\citenamefont {Gross}(1961)}]{Gross1961}%
  \BibitemOpen
  \bibfield  {author} {\bibinfo {author} {\bibfnamefont {E.~P.}\ \bibnamefont {Gross}},\ }\bibfield  {title} {\bibinfo {title} {{Structure of a quantized vortex in boson systems}},\ }\href {https://doi.org/https://doi.org/10.1007%2FBF02731494} {\bibfield  {journal} {\bibinfo  {journal} {Il Nuovo Cimento}\ }\textbf {\bibinfo {volume} {20}},\ \bibinfo {pages} {454–477} (\bibinfo {year} {1961})}\BibitemShut {NoStop}%
\bibitem [{\citenamefont {Pitaevskii}(1961)}]{Pitaevskii1961}%
  \BibitemOpen
  \bibfield  {author} {\bibinfo {author} {\bibfnamefont {L.}~\bibnamefont {Pitaevskii}},\ }\bibfield  {title} {\bibinfo {title} {{Vortex Lines in an Imperfect Bose Gas}},\ }\href {http://www.jetp.ras.ru/cgi-bin/e/index/r/40/2/p646?a=list} {\bibfield  {journal} {\bibinfo  {journal} {J. Exp. Theor. Phys.}\ }\textbf {\bibinfo {volume} {13}},\ \bibinfo {pages} {646} (\bibinfo {year} {1961})}\BibitemShut {NoStop}%
\bibitem [{Note1()}]{Note1}%
  \BibitemOpen
  \bibinfo {note} {(Supplemental Material) Details of our numerical settings and movies of our simulations can be accessed online.}\BibitemShut {Stop}%
\bibitem [{\citenamefont {Sirovich}(1967)}]{Sirovich1987Snapshot}%
  \BibitemOpen
  \bibfield  {author} {\bibinfo {author} {\bibfnamefont {L.}~\bibnamefont {Sirovich}},\ }\bibfield  {title} {\bibinfo {title} {{Turbulence and the dynamics of coherent structures, Parts I-III.}},\ }\href {https://www.jstor.org/stable/43637457} {\bibfield  {journal} {\bibinfo  {journal} {Q. Appl. Math.}\ }\textbf {\bibinfo {volume} {45}},\ \bibinfo {pages} {561} (\bibinfo {year} {1967})}\BibitemShut {NoStop}%
\bibitem [{Note2()}]{Note2}%
  \BibitemOpen
  \bibinfo {note} {(Supplemental Material) Detailed settings of our POD analysis can be accessed online.}\BibitemShut {Stop}%
\bibitem [{Note3()}]{Note3}%
  \BibitemOpen
  \bibinfo {note} {(Supplemental Material) Videos of the time development of vorticity and density can be accessed online.}\BibitemShut {Stop}%
\bibitem [{\citenamefont {von Kármán}\ and\ \citenamefont {Rubach}(1912)}]{Karman1912KVS}%
  \BibitemOpen
  \bibfield  {author} {\bibinfo {author} {\bibfnamefont {T.}~\bibnamefont {von Kármán}}\ and\ \bibinfo {author} {\bibfnamefont {H.~L.}\ \bibnamefont {Rubach}},\ }\bibfield  {title} {\bibinfo {title} {{The Mechanism of Fluid Resistance}},\ }\href@noop {} {\bibfield  {journal} {\bibinfo  {journal} {Physik Z.}\ }\textbf {\bibinfo {volume} {13}},\ \bibinfo {pages} {49} (\bibinfo {year} {1912})}\BibitemShut {NoStop}%
\bibitem [{\citenamefont {Khademinezhad}\ \emph {et~al.}(2015)\citenamefont {Khademinezhad}, \citenamefont {Talebizadeh},\ and\ \citenamefont {Rahimzadeh}}]{Taha2015square}%
  \BibitemOpen
  \bibfield  {author} {\bibinfo {author} {\bibfnamefont {T.}~\bibnamefont {Khademinezhad}}, \bibinfo {author} {\bibfnamefont {P.}~\bibnamefont {Talebizadeh}},\ and\ \bibinfo {author} {\bibfnamefont {H.}~\bibnamefont {Rahimzadeh}},\ }\bibfield  {title} {\bibinfo {title} {{Numerical Study of Unsteady Flow around a Square Cylinder in Compare with Circular Cylinder}},\ }\href {https://www.researchgate.net/publication/272416018} {\bibfield  {journal} {\bibinfo  {journal} {Conference: 1st National Conference on Fluid Flow, Heat and Mass TransferAt: Tehran, Iran}\ }\textbf {\bibinfo {volume} {1}} (\bibinfo {year} {2015})}\BibitemShut {NoStop}%
\bibitem [{Note4()}]{Note4}%
  \BibitemOpen
  \bibinfo {note} {(Supplemental Material) Details on the way to compute $l/d$ and $\protect \mathit {St}$ can be accessed online.}\BibitemShut {Stop}%
\bibitem [{\citenamefont {Christenhusz}\ \emph {et~al.}(2024)\citenamefont {Christenhusz}, \citenamefont {Safavi-Naini}, \citenamefont {Rubinsztein-Dunlop}, \citenamefont {Neely},\ and\ \citenamefont {Reeves}}]{Christenhusz2024Rs}%
  \BibitemOpen
  \bibfield  {author} {\bibinfo {author} {\bibfnamefont {M.~T.~M.}\ \bibnamefont {Christenhusz}}, \bibinfo {author} {\bibfnamefont {A.}~\bibnamefont {Safavi-Naini}}, \bibinfo {author} {\bibfnamefont {H.}~\bibnamefont {Rubinsztein-Dunlop}}, \bibinfo {author} {\bibfnamefont {T.~W.}\ \bibnamefont {Neely}},\ and\ \bibinfo {author} {\bibfnamefont {M.~T.}\ \bibnamefont {Reeves}},\ }\href {https://arxiv.org/abs/2406.14049} {\bibinfo {title} {{}}} (\bibinfo {year} {2024}),\ \Eprint {https://arxiv.org/abs/2406.14049} {arXiv:2406.14049} \BibitemShut {NoStop}%
\end{thebibliography}%

\onecolumngrid
\newpage

\section*{Supplemental material}
Movies showing the time evolution of the wake presented in the main text is available from 
\URL{\href{https://www.youtube.com/playlist?list=PLtG3C55s-pk4Wn5XD7K-yohft-S6DI_Tm}{this https URL}}
.

\section{\label{app:settigs}Numerical Settings}
We consider two-dimensional flow around a square prism hardwall potential of cross-section $d\times d$ moving with a fixed speed $U$ in the $x$ direction in a superfluid at rest in the laboratory frame.
The size of the obstacle is varied in the range $d=12\xi\sim 64\xi$, based on the healing length $\xi=\hbar/\sqrt{m\mu}$, which characterizes the core size of a quantum vortex.
Here, $\hbar$ is the reduced Planck constant, $m$ is the mass of Bose particles constituting the superfluid, and $\mu$ is the chemical potential.
To form a turbulent wake, we perform simulations for $U = 0.6c \sim 0.9c$, which is sufficiently larger than the critical velocity for quantum vortex nucleation.
Here, $c$ is the sound speed.
The system size varies with the obstacle size and is set to $15d \times 8d$.

\begin{figure}[htbp]
\centering
\includegraphics[width=7.5cm]{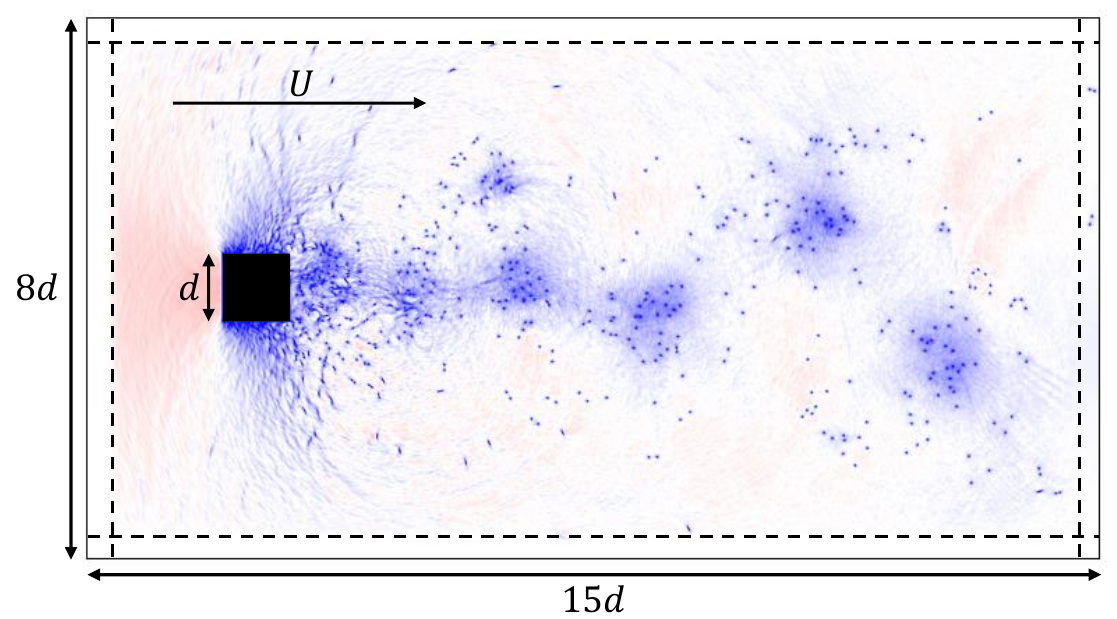}
\caption{
Illustration of the system.
A square prism hard-core potential with a cross-sectional area of $d \times d$ moves with a fixed speed $U$ in the $x$-direction.
The system size is $15d \times 8d$, and Neumann boundary conditions are imposed on the boundaries.
Note that dissipation is added near the boundaries to eliminate quantum vortices.
}
\label{fig:system}
\end{figure}

We solve a modified GP equation in a coordinate system in which the object is at rest.
The equations are nondimensionalized using the length scale $\xi$ and the energy scale $\mu$ for the calculations.
The computational grid size is set to $\Delta x=\Delta y=0.5\xi$.
A dissipation term is added to this equation to obtain a non-equilibrium steady state for the superfluid wake.
\begin{equation}
    \partial_t\psi=\frac{1}{i\hbar}\left[ H+i\hbar U\partial_x \right]\psi-\gamma(\bm{x})H\psi.
\end{equation}
with the operator
\begin{equation}
    H= -\frac{\hbar^2 \bm{\nabla}^2}{2m}-\mu +gn+V_{\rm ext}
\end{equation}
Here, $g$ is the coupling constant proportional to the s-wave scattering length of the atomic interaction and $n$ is the fluid density.
The external potential $V_{\rm ext}(\bm{x})$ represents the square obstacle whose interior completely excludes fluid as schematically illustrated as a black square in Fig.~\ref{fig:system}.
The dissipation coefficient $\gamma(\bm{x})$ in the third term on the right hand side is defined as follows;
\begin{equation}
    \begin{aligned}
    \gamma(\bm{r})=1&-0.5\left\{1-\tanh\left[\frac{|x|-(L_x/2-L_\gamma)}{w_\gamma}\right]\right\}
    \cdot0.5\left\{1-\tanh\left[\frac{|y|-(L_y/2-L_\gamma)}{w_\gamma}\right]\right\}
    \end{aligned}
\end{equation}
The parameters determining the size of the dissipation region are set to $L_\gamma=18\xi, w_\gamma=7\xi$.
As a result, $\gamma(\bm{x})$ becomes effective only near the boundaries.
This method is similar to the one implemented by Reeves et al. \cite{Reeves2015Rs}.
However, in contrast to previous studies, the Neumann boundary condition is imposed on the boundaries of the system.
In our method, quantum vortices that reach the boundary vanish on the spot.

\newpage

\section{\label{app:POD}Process of POD}
We use Snapshot POD, proposed by Sirovich \cite{Sirovich1987Snapshot}, one of the methods of Proper Orthogonal Decomposition (POD).
This method is that allows for obtaining POD bases with lower computational cost when the spatial resolution is high after discretization.
The deviation of each time-series data (Snapshot) from its mean is treated as a single vector, and the following eigenvalue problem is formulated for the matrix $\bm{X}$ composed of these vectors.
Treating each of the $n$ Snapshot as a vector, the matrix $\bm{X}$ is constructed.
\begin{equation}
    \bm{X}=\left[ \sqrt{\omega_1}\bm{x}_1, \sqrt{\omega_2}\bm{x}_2, \cdots \sqrt{\omega_n}\bm{x}_n \right]
    \label{eq:PODmat}
\end{equation}
$\bm{x}_m$ is obtained by subtracting the time mean from the Snapshot. $\omega_m$ represents the weighting of each data point, and in this study, where snapshots are taken at equal intervals, it is simply given by $\omega_m=1/n$.
\begin{equation}
    \bm{X}^T\bm{X}\bm{u}_k=\lambda_k\bm{u}_k
    \label{eq:POD}
\end{equation}
The following operations are performed on the obtained eigenvectors $\bm{u}_k$:
\begin{equation}
    \bm{\varphi}_k=\frac{\bm{X}\bm{u}_k}{\sqrt{\lambda_k}}
    \label{eq:ev}
\end{equation}
By performing the reverse operation of vectorizing the time-series data on the obtained $\bm{\varphi}_k$, and visualizing as a numerical distribution, the spacial patterns shown in the main text can be obtained.
The larger the corresponding eigenvalue $\lambda_k$, the higher the contribution to the original flow, and they are called mode 1, 2... in that order.
The contribution of each mode can be calculated by dividing $\lambda_k$ by the sum of the eigenvalues of all modes.
Also, by taking the inner product of $\bm{\varphi}_k$ and $\bm{x}_m$, the extent to which the POD basis components are present in the data can be calculated.
By performing this operation on all time-series data, it is possible to determine the periodicity of each mode's appearance.

In the actual calculation, the system is first evolved for a relaxation time $t_\textrm{relax}$, and snapshots are then taken from the resulting nonequilibrium steady state.
Assuming that the relaxation time is proportional to $d/U$, we set $t_\textrm{relax} = 32d/U$, which is sufficiently long to achieve a nonequilibrium steady state.
As shown in \cite{Reeves2015Rs}, the time evolution of a superfluid wake has a characteristic periodic time
\begin{equation}
    \tau=f^{-1}=\frac{d}{\mathit{St} U}, 
    \label{Eq_tau}
\end{equation}
determined by the Strouhal number $\mathit{St}$.
To obtain statistically reliable results, time evolution data up to more than $40\tau$ after $t=t_\textrm{relax}$ were used for our POD analysis.
Since $\mathit{St}$ in superfluid wakes is nontrivial, we set it to $\mathit{St} = 0.16$, which is close to the Strouhal number obtained for a square cylinder in classical fluids.

To capture the microscopic motion of quantum vortices in a vortex bundle using POD, the snapshot interval must be set to be comparable to or shorter than the system’s microscopic time scale $\tau_m$.
Here, we estimate the microscopic time scale as $\tau_m = \sigma_b^2 / \kappa$ using the average spacing $\sigma_b$ between neighboring vortices in a single vortex bundle.
Since we defined $\sigma_b \equiv \sqrt{\frac{S_b}{N_v}}$ in the main text, using the expression for $N_v$, $\tau_m$ becomes:
\begin{equation}
    \tau_m=\frac{\sigma_b^2}{\kappa}=\frac{S_b}{N_v\kappa}=\frac{S_b}{\kappa}
    \left[2\sqrt{2}\frac{l}{d}\left( 1-\frac{l}{d}\mathit{St} \right) Re_s \right]^{-1}
    \label{tau_m1}
\end{equation}
As with the estimation of $\sigma_b$ in the main text, we assume $S_b \sim d^2$.
Given that $Re_s = dU / \kappa$, $\tau_m$ becomes:
\begin{equation}
    \tau_m=\left[2\sqrt{2}\frac{l}{d}\left( 1-\frac{l}{d}\mathit{St} \right) \right]^{-1}\frac{d}{U}
    \label{tau_m2}
\end{equation}
The ratio between the macroscopic time scale $\tau$ and the microscopic time scale $\tau_m$ is given as follows:
\begin{equation}
    \frac{\tau}{\tau_m}=2\sqrt{2}\frac{l}{d}\left( \frac{1}{\mathit{St}}-\frac{l}{d} \right)
    \label{macro/micro}
\end{equation}
Using the assumptions from the main text, $d/l \approx b/l = \cosh^{-1}(\sqrt{2})/\pi \approx 0.2806$ and $\mathit{St} = 0.15$, we obtain $\tau / \tau_m \sim 31.3$.
Therefore, it can be inferred that acquiring approximately 30 to 40 snapshots per $\tau$ is sufficient.
In this study, 40 snapshots are taken per $\tau$.
The total number of snapshots is $n = 40 \times 40 = 1600$.

In order to reflect the microscopic motion of each quantum vortex,
it is necessary to take snapshots with a sufficiently small time resolution.
Therefore, we used 1600 snapshots, which is considerably more than the usual POD analysis.

\newpage

\section{\label{app:Nv}Method for Calculating the spatial and temporal periods from POD}
From the POD results, the spatial and temporal periods of the quasi-classical Kármán vortex street can be estimated, allowing the calculation of $l/d$ and $\mathit{St}$.

The dominant modes obtained from the vorticity POD with Gaussian blur using $\sigma = \sigma_b$ exhibit symmetric patterns.
By extracting the numerical distribution at $y=0$ from this pattern, the waveform can be obtained, and its spatial period can be used to calculate $l$.

\begin{figure}[htbp]
\centering
\includegraphics[width=15cm]{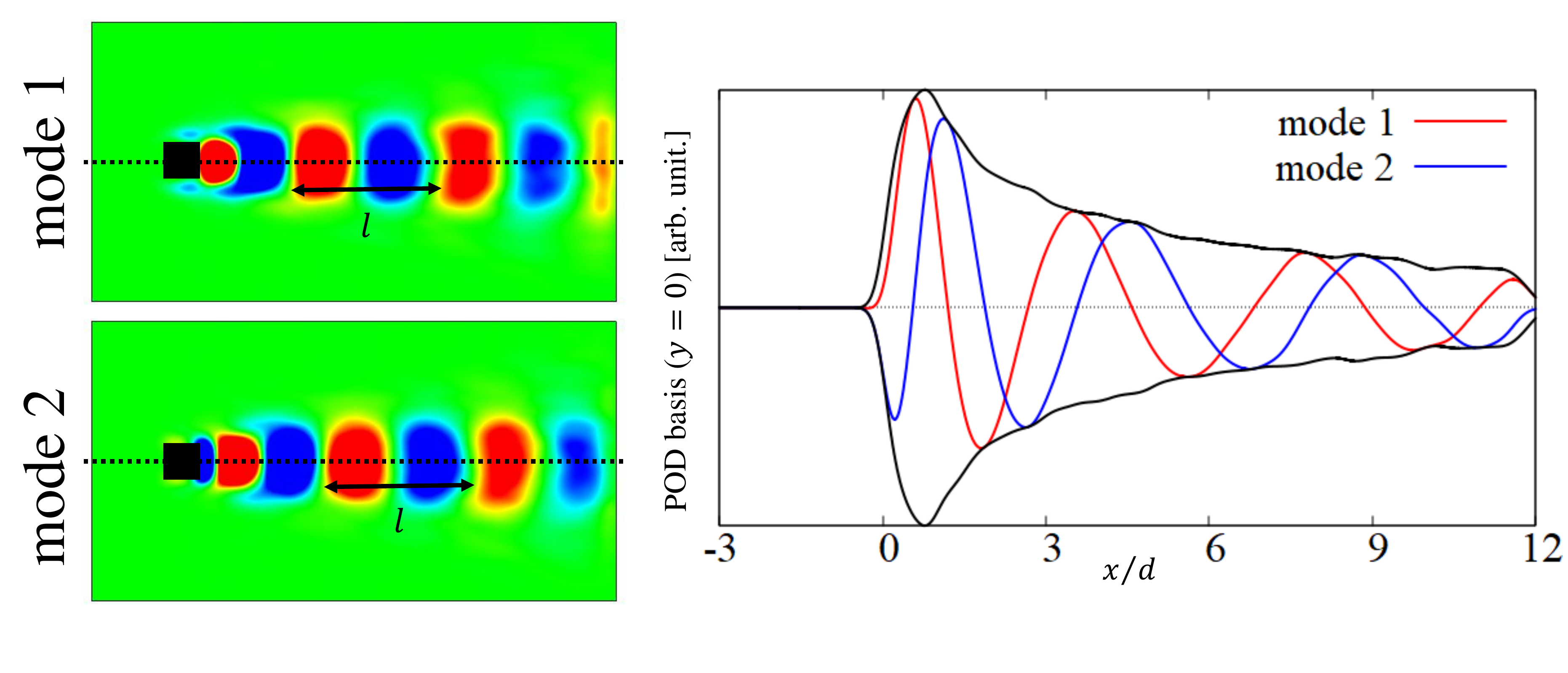}
\caption{
The pattern corresponding to the dominant mode and its numerical distribution at $y = 0$ in the case of $d = 64\xi$, $U = 0.7c$.
The trailing edge of the obstacle is set as $x=0$.
The spatial period is considered to correspond to the streamwise spacing $l$ between vortex bundles in the quasi-classical Kármán vortex street.
By using Gaussian smoothing with $\sigma = \sigma_b$, the waveform becomes smoother, making it easier to read its spatial period.
}
\label{fig:l}
\end{figure}

The paired waveforms of mode 1 and mode 2 have the same period and a half-phase shift.
In this study, to reduce the influence of boundary effects, waveforms are obtained within the range of $2d<x<8d$, and $l$ is estimated.

By taking the inner product of the POD basis corresponding to the dominant mode, $\bm{\varphi}_1$ and $\bm{\varphi}_2$, with each time-series data point, $\bm{x}_m$, and arranging them in order, a waveform representing the occurrence cycle of the dominant mode is obtained [see II.].
This temporal frequency corresponds to the frequency $f$ of the quasi-classical Kármán vortex street.
Using the value of $f$ obtained in this way, the Strouhal number corresponding to the quasi-classical Kármán vortex street can be calculated from POD as $\mathit{St} = fd/U$.

\begin{figure}[htbp]
\centering
\includegraphics[width=15cm]{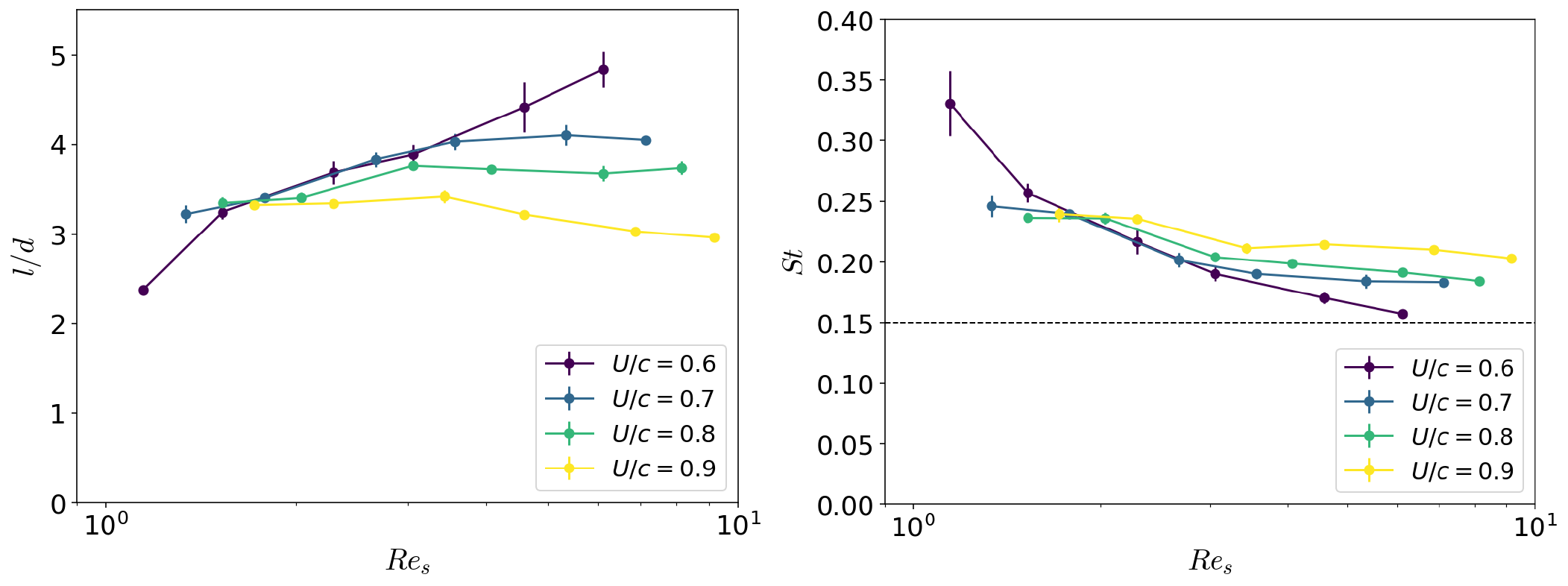}
\caption{
$l/d$ (left) and $\mathit{St}$ (right) calculated from POD for each parameter.
In the $\mathit{St}$ plot, the Strouhal number for the wake behind a square prism in classical fluids, $\mathit{St} = 0.15$, is indicated by a dashed line.
}
\label{fig:lSt}
\end{figure}

The graph of $l/d$ and $\mathit{St}$ as functions of $Re_s$ calculated in this study is shown in Fig.~\ref{fig:lSt} (An enlarged and revised version of the figure shown in the main text).
By substituting these values into the expression for $N_v$ presented in the main text, the number of quantum vortices contained in a vortex bundle can be estimated using only the information obtained from POD.

\newpage

\section{\label{app:PODresults}POD results of each parameters}
\begin{figure}[htbp]
\centering
\includegraphics[width=11cm]{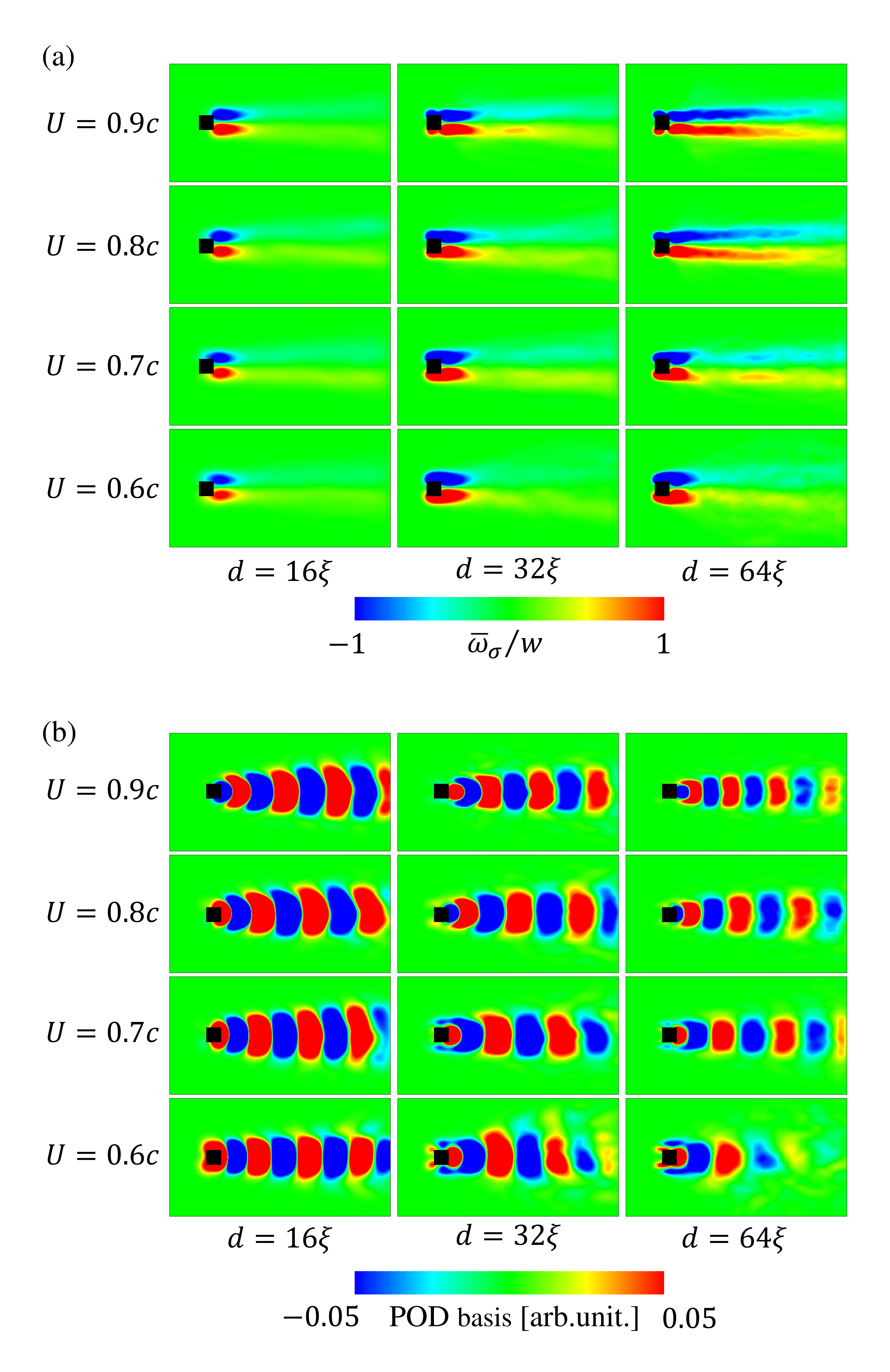}
\caption{
Results of POD using the Gaussian blur with $\sigma=\sigma_b$ for each parameters.
(a) Normalized mean vorticity. The mean filtered vorticity $\bar{\omega}_\sigma$ is rescaled by $w=\xi^2\kappa/2\pi \sigma^2 \Delta x^2$.
(b) First contributing mode.
}
\label{fig:PODresults}
\end{figure}

\end{document}